\documentstyle[aps,prb,preprint,tighten]{revtex}

\newcommand{\br}{{\bf r}}
\newcommand{\bk}{{\bf k}}
\newcommand{\bg}{{\bf g}}
\newcommand{\bxi}{\hbox{\boldmath $\xi$}}
\newcommand{\bRt}{\bf{\tilde R}}
\newcommand{\bgt}{\bf{\tilde g}}
\newcommand{\lbar}{\bar l}
\newcommand{\mbar}{\bar m}
\newcommand{\be}{\begin{equation}}
\newcommand{\ee}{\end{equation}}
\newcommand{\bea}{\begin{eqnarray}}
\newcommand{\eea}{\end{eqnarray}}

\begin{document}
\draft

\title{Multi-band energy spectra of
spin-1/2 electrons with two-dimensional magnetic modulations}

\author {Ming-Che Chang$^1$ and Min-Fong Yang$^2$}
\address{$^1$Department of Physics, National Tsing Hua University,
Hsinchu, Taiwan\\
$^2$Center of General Education, Chang Gung University,
Kweishan, Taoyuan, Taiwan}

\date{\today}

\maketitle

\begin{abstract}
The energy spectra of spin-1/2 electrons under two-dimensional
magnetic field modulations are calculated beyond the one-band
approximation.  Our formulation is generally applicable to a
modulation field with a rectangular lattice symmetry.  The field
distribution within a plaquette is otherwise arbitrary.  
The spectra being obtained are 
qualitatively different from their electric modulated 
counterparts.
Peculiar features of the spectra are that, for an electron with a $g$
factor precisely being equal to two, no matter how strong the
modulation is, the zero-energy level seems to be unaffected by the
modulation and is separated from higher energy levels with a nonzero
energy gap.  Moreover, there is a two-fold degenerancy for all states
with positive energies with respect to spin flip.  These features agree
with earlier analytical studies of the periodically magnetic modulated 
systems. 
\end{abstract}

\narrowtext

\section{Introduction}
Due to the progress of the submicron technology, one begins to observe
quantum behaviors in the transport measurements of the field modulated
two-dimensional electron gas (2DEG).
Much effort has been devoted to the study of the spectral and the
transport properties of a 2DEG in a periodic {\it magnetic} field with
a nonzero uniform background $B_0$.  These studies can be divided into
two classes, depending on whether the field modulation is
one dimensional
\cite{1dmm,1dmmc,1dmmi,1dmms,1dmme,1dmm2dem,1dmmspin,carmona,ye} or
two dimensional.  \cite{2dmmc,2dmm2dem,2dmm,chang94,dubrovin,rom}

For a one-dimensional magnetic modulated (1DMM) system, there are two
characteristic length scales:  the magnetic length,
$\lambda=\sqrt{\hbar/eB_0}$, associated with the uniform background
$B_0$ and the period $a$ of modulation.  By varying the ratio
$\lambda/a$, the electron mobility and the magnetoresistance oscillate
between extrema.  \cite{1dmm} The oscillating behavior of the latter
is similar to the Weiss oscillation in a one-dimensional electric
modulated (1DEM) 2DEG.\cite{1dem} These oscillations manifest the 
variation of the band widths:  the (longitudinal) conductivity is 
proportional 
to the width of the Landau-level (LL) broadening, which is an oscillating
function of both $\lambda/a$ and LL indices due to the field modulation.
Under special conditions (the so-called flat band conditions), the
band width can be zero and electrons become immobile.  Besides the
transport property, other aspects of the 1DMM system have also been
studied, such as collective excitations,\cite{1dmmc} inelastic light
scatterings,\cite{1dmmi} surface states,\cite{1dmms} effect of
electron-electron interactions,\cite{1dmme} and the effect due to an
additional two-dimensional electric modulation.\cite{1dmm2dem} The
1DMM systems may also be used as spin polarizers for magnetic
dipoles.\cite{1dmmspin} Recently, the 1DMM systems have been realized
experimentally by covering a regular array of
superconductor\cite{carmona} or micromagnet\cite{ye} on the top of a
2DEG, in which the observed magnetoresistance oscillation agrees very
well with the theoretical prediction.

For the two-dimensional magnetic modulated (2DMM) systems, the Landau
levels are not only broadened but also split to several subbands with
an intricate fractal structure.  \cite{2dmmc,2dmm2dem} This is similar
to the Hofstadter spectra for the two-dimensional electric modulated
(2DEM) systems.\cite{2dem} In fact, within the one-band approximation,
both spectra are exactly the same to linear order in modulation
fields.  \cite{2dmm2dem,2dmm} For a square lattice in the one-band
approximation, a Landau band is split to $p$ subbands when there are
$p/q$ ($p$ and $q$ are relative prime integers) flux quanta per
plaquette.  Previous calculations of the 2DMM systems are either
restricted to the one-band approximation, \cite{2dmm2dem} or to the
multiband calculation but with a specific magnetic field such that the
flux quantum per plaquette is one (or one half).\cite{2dmm,chang94}
Other aspects of the 2DMM systems have been studied, such as the
collective excitations \cite{2dmmc}
%which also show a Hofstadter-like fractal structure,
and the degeneracy of the ground states.\cite{dubrovin,rom} There are
some recent attempts to observe the peculiar transport properties due
to the fractal band structure.  However, as far as we know, this goal
has not been achieved for the 2DMM systems.\cite{2dmmtry}

In the present work, the energy spectra of a spin-1/2 2DEG under
two-dimensional magnetic field modulations are calculated {\it beyond
the one-band approximation}, in which the Zeeman term is also included.
In most of the above studies, the Zeeman effect is not included,
however.  For the electric modulated systems, neglect of the Zeeman
term is justified, because the periodic electric field does not couple
to electron spin and this term only contributes to a constant energy
shift, $(g_e e \hbar/4m)B_0\sigma_z$, where $g_e$ is the electron $g$
factor and $\sigma_z$ is $+1(-1)$ for spin-up (spin-down) electrons.
However, this is not the case for the magnetic modulated systems.
After deriving the multi-band Harper equation, which is generally
applicable to magnetic modulations with {\it arbitrary} strength and
shape, as long as the field has a rectangular lattice symmetry, we
show that the inclusion of the Zeeman term leads to qualitative
changes in the energy spectra.  Particularly, when $g_e=2$, the most
disparate result occurs for the lowest energy level --- it is {\it
not} broadened by the field modulation, and is separated from higher
energy bands by a finite gap. Moreover, there exists a two-fold
degenerancy for all states with positive energies with respect to spin
flip.  These results agree with earlier
mathematical analysis of the 2DMM systems.\cite{dubrovin,klauder}

The paper is organized as follows:  the multi-band formalism is
presented in Sec.~II; the band structure is presented in Sec.~III; and
Sec.~IV is devoted to summary and discussions.

\section{Multi-band formalism}

\subsection{Magnetic translation symmetry}

We consider a 2DEG under the influence of a magnetic modulation with a
rectangular symmetry.  The Hamiltonian is
\be
H=\frac{1}{2}\left(-i\frac{\partial}{\partial \br}+{\bf A}_0(\br)+{\bf
a}(\br)\right)^2+\frac{g_e}{4}B(\br)\sigma_z,
\ee
where ${\bf A}_0(\br)$ and ${\bf a}(\br)$ are the vector potentials for the
uniform background field $B_0$ and the modulation field
$b(\br)=B(\br)-B_0$, respectively.  In this paper, unless specified
explicitly, we choose $\lambda=\sqrt{\hbar/eB_0}$ as the unit of
length, $\hbar\omega_c$ as the unit of energy ($\omega_c=eB_0/m$ is
the cyclotron frequency), and $B_0$ as the unit of magnetic field.  In
the absence of modulation, the Hamiltonian $H_0$ can be solved exactly
with eigenvalues $E^{(0)}_n=n+1/2+g_e\sigma_z/4$.  \cite{Landau} $H$
can be expanded as $H_0+H_1+H_2+H_\sigma$, where $H_1$ and $H_2$ are
the terms linear and quadratic in the vector potential ${\bf a}(\br)$,
respectively, and $H_\sigma =(g_e/4)b(\br)\sigma_z$ is the {\it
modulated} Zeeman term.  The vector potential ${\bf a}(\br)$ can be
Fourier decomposed as ${\bf a}(\br)=\sum_{\bg \neq 0} {\bf a}_\bg
e^{i\bg\cdot\br}$, where $\bg$ are the reciprocal lattice vectors of
the rectangular lattice.  By choosing the Coulomb gauge, the Fourier
components ${\bf a}_\bg$ are equivalent to $ib_\bg \bg\times{\bf\hat
z}/g^2$, where $b_\bg$ are the Fourier components of $b(\br)$ and
$g=|\bg|$ (not to be confused with the electron $g$ factor $g_e$).  It
is convenient to rewrite the exponential $e^{i\bg\cdot\br}$ as
$e^{i\bg\cdot\bxi}e^{i\bg\cdot{\bf R}}$, where the electron coordinate
$\br$ is decomposed into a fast-moving cyclotron coordinate $\bxi$ and
a slow-moving guiding-center coordinate ${\bf R}=\br-\bxi$.  (See
Appendix \ref{guiding}.) Then it can be shown that
\be
H_1=-\sum_{\bg \neq 0} \frac{b_\bg}{g^2}\frac{\partial
e^{i\bg\cdot{\bxi}\lambda}}{\partial\lambda}
|_{\lambda=1}e^{i\bg\cdot{\bf R}}.
\label{HH1}
\ee
Similarly,
\be
H_2=-\frac{1}{2}\sum_{\bg \neq 0} \sum_{\bg' \neq 0}
\bg\cdot\bg'\frac{b_\bg}{g^2}\frac{b_{\bg'}}{g'^2}e^{i(\bg+\bg')\cdot\bxi}
e^{i(\bg+\bg')\cdot{\bf R}},
\label{HH2}
\ee
and
\be
H_\sigma=\frac{g_e \sigma_z}{4}\sum_{\bg \neq 0}
b_\bg e^{i\bg\cdot\bxi} e^{i\bg\cdot{\bf R}}.
\label{HHs}
\ee

Due to the underlying magnetic translation symmetry \cite{zak} of the
Hamiltonian, it is convenient to diagonalize the Hamiltonian on a
basis which respects this symmetry.  The unperturbative basis can be
constructed as follows.  By choosing a Landau gauge with ${\bf
A}_0(\br)=(-y,0)$, the magnetic translation operators are
\be
T_1=e^{a_1\partial/\partial x},\ \
T_2=e^{a_2(\partial/\partial y-ix)},
\label{T}
\ee
where $a_1$ and $a_2$ are the lattice constants for the rectangular
lattice.  It is not difficult to show that, if there are $p/q$ flux
quanta per plaquette with an area $a_1a_2$, then $T_1$, $T_2^q$, and
$H$ mutually commute.  This is also true for the unmodulated
Hamiltonian $H_0$, of course.
Thus we can construct the explicit form of the magnetic Bloch states
for $H_0$, which are the common eigenstates of $H_0$, $T_1$ and
$T_2^q$:  \cite{chang94}
\be
|n,\bk\rangle=\sum_{l=-\infty}^{\infty} {\bar d}_l e^{-i(q/p)k_2a_2l}
|n,k_1-\frac{2\pi}{a_1}l\rangle,
\label{n}
\ee
where $n$ is the LL index, $\bk=(k_1, k_2)$ is the magnetic Bloch
momentum, ${\bar d}_l$ are complex coefficients that are periodic in
$l$ with period $p$ (i.e., ${\bar d}_{l+p}={\bar d}_l$), and
$|n,k_1\rangle$ are the common eigenstates of $H_0$ and $T_1$.

Since both $H$ and $H_0$ are diagonal with respect to $\bk$, it is
clear that for the modulation part $H'=H-H_0$, we have
$\langle n,\bk|H'|n',\bk'\rangle=\langle
n,\bk|H'|n',\bk\rangle\delta_{\bk,\bk'}$.  Therefore, the $\alpha$-th
eigenstate of $H$ can be written as
\be
|\alpha,\bk\rangle
=\sum_{n=0}^\infty\sum_{l=-\infty}^{\infty}
d_{n,l}^{(\alpha)} e^{-i(q/p)k_2a_2l} |n,k_1-\frac{2\pi}{a_1}l\rangle,
\label{E}
\ee
where the unknown coefficients $d_{n,l}^{(\alpha)}$, as ${\bar d}_l$
above, are periodic in $l$ with period $p$.  Basically, the strategy
below is to diagonalize the Hamiltonian matrix on the unperturbed
basis and to solve for its eigenvalues.

In deriving the matrix elements of the Hamiltonians in Eqs.~(2)--(4)
on the unperturbed basis, the expression
$\langle n,k_1-2\pi l/a_1|e^{i\bg\cdot\bxi}e^{i\bg\cdot{\bf R}}
|n',k_1-2\pi l'/a_1\rangle$ will be encountered frequently,
thus we focus on its derivation first.  Firstly we rewrite the
exponential in a slightly different form, $e^{i\bg\cdot\bxi}
e^{i\bgt\cdot\bRt}$, to connect with the magnetic translation
symmetry, where $\bgt=\bg\times{\hat z}$ and ${\bf\tilde R}={\bf
R}\times{\bf\hat z}$.  Since the two dynamical
variables, $\bxi$ and $\bRt$, decouple and operate on different parts
of the Hilbert space (see Appendix \ref{guiding}), the matrix elements
of $e^{i\bg\cdot\bxi}e^{i\bgt\cdot\bRt}$ can be evaluated with the
help of Eq.~(\ref{R}).  The result is
\be
\left \langle
n,k_1-\frac{2\pi}{a_1}l\left|e^{i\bg\cdot\bxi}e^{i\bgt\cdot\bRt}
\right|n',k_1-\frac{2\pi}{a_1}l'\right\rangle =
\delta_{l,l'-\lbar}P_{k_1 l}(\bg)U_{nn'}(\bg),
\label{basic}
\ee
where $\bg=(g_1, g_2)=(2\pi \lbar/a_1,2\pi \mbar/a_2)$,
$P_{k_1 l}(\bg)=e^{-\pi i\lbar\mbar
q/p}e^{2\pi ik_1\mbar/a_2} e^{-2\pi i\mbar l q/p}$, and $U_{nn'}(\bg)=\langle
n|e^{i\bg\cdot \bxi}|n'\rangle$.  The magnetic flux condition, $a_1a_2=2\pi
p/q$, has been used.

\subsection{Multi-band Harper equation}

When the energy eigenstate $|\alpha,\bk\rangle$ is expanded on the
basis of $|n,k_1-2\pi l/a_1\rangle$ using Eq.~(\ref{E}), the
eigenvalue equation,
$H|\alpha,\bk\rangle=E_\alpha(\bk)|\alpha,\bk\rangle$, takes the
following form,
\be
E^{(0)}_n d_{n,s}^{(\alpha)} + \sum_{n',l'} e^{-i(q/p) k_2a_2(l'-l)}
\langle n,k_1-\frac{2\pi l}{a_1}|H'|n',k_1-\frac{2\pi l'}{a_1}\rangle
d_{n',s'}^{(\alpha)}=E_\alpha d_{n,s}^{(\alpha)},
\label{Harper1}
\ee
where $l=pr+s$ and $l'=pr'+s'$.  ($r, s, r',$ and $s'$ are all
integers such that $0\leq s,s' < p$.) Firstly we need to calculate the
matrix elements of $H'$.  With the help of
Eqs.~(\ref{HH1})--(\ref{HHs}) and (\ref{basic}), we have
\bea
\left\langle n,k_1-\frac{2\pi l}{a_1}|H_1|n',k_1-\frac{2\pi
l'}{a_1}\right\rangle
&=& -\delta_{l,l'-\lbar}\sum_{\bg \neq 0}\frac{b_\bg}{g^2}
P_{k_1 s}(\bg)\frac{\partial
U_{nn'}(\bg\lambda)}{\partial\lambda}|_{\lambda=1},\cr
\left\langle n,k_1-\frac{2\pi l}{a_1}|H_2|n',k_1-\frac{2\pi
l'}{a_1}\right\rangle
&=& -\frac{1}{2}\delta_{l,l'-\lbar-\lbar'}\sum_{\bg \neq 0}
\sum_{\bg' \neq 0}
\bg\cdot\bg'\frac{b_\bg}{g^2}\frac{b_{\bg'}}{g'^2}
P_{k_1 s}(\bg+\bg')U_{nn'}(\bg+\bg'),\cr
\left\langle n,k_1-\frac{2\pi l}{a_1}| H_\sigma
|n',k_1-\frac{2\pi l'}{a_1}\right\rangle
&=& \frac{g_e \sigma_z}{4} \delta_{l,l'-\lbar}
\sum_{\bg \neq 0} b_\bg P_{k_1 s}(\bg)U_{nn'}(\bg).
\label{long}
\eea

Combining Eqs.~(\ref{Harper1}) and (\ref{long}), we finally obtain
\bea
E^{(0)}_n d_{n,s}^{(\alpha)} &+& \sum_{n'}\sum_{\bg \neq 0} b_\bg
{\cal P}_{\bk s}(\bg)
\left[-\frac{1}{g^2}\frac{\partial
U_{nn'}(\bg\lambda)}{\partial\lambda}|_{\lambda=1}+\frac{g_e \sigma_z}{4}
U_{nn'}(\bg)\right] d_{n',s+{\bar s}}^{(\alpha)} \cr
&-&\frac{1}{2}\sum_{n'}\sum_{\bg \neq 0} \sum_{\bg' \neq 0} \bg\cdot\bg'
\frac{b_\bg}{g^2}\frac{b_{{\bf g}'}}{g'^2}{\cal P}_{\bk s}(\bg+\bg')
U_{nn'}(\bg+\bg') d_{n',s+{\bar s}+{\bar s}'}^{(\alpha)}
=E_\alpha d_{n,s}^{(\alpha)},
\label{Harper2}
\eea
where ${\cal P}_{\bk s}(\bg)=e^{-2\pi ik_2\lbar/a_1}
P_{k_1 s}(\bg)=e^{i\bk\cdot(\bg\times{\bf\hat
z})}e^{-ig_1g_2/2}e^{-i g_2 (2\pi s/a_1)} $.
It is a multi-band generalization of the Harper equation \cite{2dem}
(see Eq.~(\ref{Harper})), thus Eq.~(\ref{Harper2}) is called as the
multi-band Harper equation.  \cite{petschel} It has to be solved in
conjunction with the following identities concering inter-LL
transitions (for $n\geq n'$),
\bea
U_{nn'}(\bg)
&=&
\sqrt{\frac{n'!}{n!}}\left(\frac{g_-}{\sqrt{2}}\right)^{n-n'}
e^{-g^2/4}L_{n'}^{n-n'},\cr
\frac{\partial
U_{nn'}(\bg\lambda)}{\partial\lambda}|_{\lambda=1}
&=&
\sqrt{\frac{n'!}{n!}}\left(\frac{g_-}{\sqrt{2}}\right)^{n-n'}
e^{-g^2/4}\left[
(n-n'-g^2/2)L_{n'}^{n-n'}-g^2L_{n'-1}^{n-n'+1}\right],
\label{inter}
\eea
where $g_-=g_1-ig_2$ and $L_{n'}^{n-n'}$ ($L_{-1}^n=0$) are the
associated Laguerre polymials with the argument $g^2/2$.

Until now, no approximation has been used.  This multi-band Harper
equation applies to general shape of magnetic field distribution with
a rectangular lattice symmetry.  To simplify the
calculation, from now on, we assume the spatial field modulation is
$b(\br)=2b_{10}[\cos(2\pi x/a)+\cos(2\pi y/a)]$ with a square lattice
symmetry (i.e., $a_1=a_2=a$). In the numerical calculations, the
eigenvalues are obtained
by diagonalizing a $(1+n_{\rm cut})p\times (1+n_{\rm cut})p$ matrix,
where the cut-off $n_{\rm cut}$ has to be large enough to ensure the
eigenvalues being obtained converge to the correct result.  The
precise spectra that include the effect of inter-LL transitions are
shown in the next section.

Before closing this subsection, we show that, under the so-called
one-band approximation, Eq.~(\ref{Harper2}) can indeed be reduced to
the usual Harper equation.  When the inter-LL transitions and the
terms quadratic in $b_\bg$ are neglected,
$d_{n,l}^{(\alpha)}\rightarrow {\bar d}_l\delta^\alpha_n$ (due to the
periodicity, ${\bar d}_{l+p}={\bar d}_l$, there are only $p$ independent
coefficients, i.e., ${\bar d}_0, \cdots, {\bar d}_{p-1}$), and then
Eq.~(\ref{Harper2}) is reduced to the one-band equation:
\be
M_n\left(\frac{q}{p}\right)\left\{ {\bar d}_{s-1} e^{2\pi ik_2/a}
+{\bar d}_{s+1} e^{-2\pi ik_2/a}
+2 {\bar d}_s \cos
\left[2\pi\left(\frac{k_1}{a}-s\frac{q}{p}\right)\right]\right\}
=\left[E_n\left(\frac{q}{p}\right)-E^{(0)}_n\right]{\bar d}_s,
\label{Harper}
\ee
where $M_n(q/p) =(b_{10}/2)[ L_n^1+L_{n-1}^1+(g_e\sigma_z/2)
L_n ] e^{-g_{10}^2/4}$ is an overall factor that scales the energy, and
$g_{10}^2=(2\pi/a)^2=2\pi q/p$.
Apart from the factor $M_n(q/p)$ where the spin-related term is included,
%Apart from the factor $M_n(q/p)$ and the spin-related term,
Eq.~(\ref{Harper}) is precisely the same as the Harper
equation for a 2DEM system.  \cite{2dem} The spectrum for
$E_n(q/p)/M_n(q/p)$
within the one-band approximation is thus trivial:  \cite{2dmm2dem,2dmm}
irrespective of the LL index $n$, it is the usual Hofstadter spectrum
calculated for a 2DEM system.  There is one exception, however.  When
$n=0$, $g_e=2$, and $\sigma_z=-1$, $M_0(q/p)$ is equal to zero, and
then $E_0(q/p)=E^{(0)}_0$, as if the field modulation exerts no influence
at all.  Actually, the equality $E_0(q/p)=E^{(0)}_0$ is valid even
beyond the one-band approximation.  This is discussed in more details
in Sec.~\ref{band}.

\section{Fractal band structure}
\label{band}

In this section we show the band structures for both weak and strong
modulations.  The influence of the Zeeman term is particularly
emphasized.
For an unmodulated 2DEG with no Zeeman effect, the energy spectrum
consists of discrete, dispersionless LLs.  \cite{Landau} These LLs are
highly degenerate because of the continuous translation symmetry,
which gives an infinite degenerancy, and the spin-flip symmetry, which
gives a two-fold degenerancy.  When a periodic modulation is
introduced, which breaks the continuous translation symmetry, we
expect that the degenerancy for each LL will be lifted.  Indeed, in
the one-band approximation, one finds that each LL is broadened and
split to several intricate energy subbands.  The way these subbands
split is the same for every LL in the one-band approximation (see the
discussion at the end of the last section).  However, when the
inter-LL transitions are included, the exact calculations shown below
reveal that the subband structures are actually different for
different LLs, thus lead to much more complicated structures.

In Fig.~1, the spectrum of electrons with $g_e=0$ under a weak square
modulation field with $b_{10}a^2=0.2\phi_0$ (in the usual units, where
$\phi_0=h/e$ is the flux quantum) is shown.  Because there is no
Zeeman splitting, there is no need to distinguish the spin-up electrons
from the spin-down electrons and the spectrum for 
only one spin direction is shown.  The calculation is done with a cut-off 
$n_{\rm cut}=9$.  The
result with a larger cut-off at $n_{\rm cut}=14$ shows no visible
difference from Fig.~1.  Notice that the abscissa is the inverse of
the magnetic flux, $q/p$.  In this and the following figures, it is
assumed that, while changing the magnetic flux by varying $B_0$, the
modulation amplitude $b_{10}$ is fixed.  For the weak modulation case,
the envelope for each energy band is largely determined by the scaling
factor $M_n(q/p)$.  Obviously, some features specific to the one-band
approximation no longer exist.  For example, the interband couplings
remove the symmetry of the butterfly diagram.  Similar effect of symmetry
breaking is also observed in the multi-band calculation for the 2DEM
systems.\cite{petschel}

The two-fold degenerancy for electron spin is lifted when the
Zeeman effect is not negligible.  In our 2DMM systems, the Zeeman
effect does not only give an energy shift, but also induce inter-LL
transitions (see Eq.~(\ref{long}) for $H_\sigma$).  Thus, the interplay
between the orbital effect ($H_1+H_2$) and the Zeeman effect
($H_\sigma$) leads to different spectral structures between the
spin-up and the spin-down electrons, which are shown in Fig.~2.  The
spectrum in Fig.~2(a) (Fig.~2(b)) is for a spin-down (spin-up)
electron with a $g$ factor equals to one.  It can be seen that, in
addition to the overall constant Zeeman energy shift $g_e\sigma_z/4$
due to the background field, the spectra show qualitative differences
from Fig.~1.  For example, the structures for the second lowest energy
bands near $q/p=0.9$ in Figs.~2(a) and (b) are visibly different from
that in Fig.~1, and are different from each other.

The spectrum in Fig.~3(a) (Fig.~3(b)) is for a spin-down (spin-up)
electron with a $g$
factor equals to two.  For this particular $g$ factor, two
significant features are observed.  The first is that the lowest energy level
for a spin-down electron is flat and equals to zero to
very high precision.  The second is that the positive-energy spectrum
is degenerate with respect to spin flip, whereas the flat band in
Fig.~3(a) has no counterpart in Fig.~3(b).  Both features
persist for stronger modulations.

The two-fold degenerancy for the positive-energy states in the case of $g_e=2$
indicates that there must be an additional symmetry even in the
presence of the Zeeman term.  It was pointed out by Aharonov and Casher
\cite{ac} that this degenerancy results from a symmetry transformation
which {\it simultaneously} changes the direction of the
electron spin and the spatial dependence of the wavefunction.  This
symmetry can be related to the supersymmetry, \cite{susy} or
to the chiral ($\gamma_5$) invariance by connecting our problem to the
($1+1$)-dimensional theory of Dirac fermions.  \cite{ac}

Moreover, based on an abstract mathematical analysis, Dubrovin and Novikov
showed that, for a spin-down electron with $g_e=2$, there always
exist the zero-energy states in the 2DMM systems, no matter how strong the
modulation is.  \cite{dubrovin} Furthermore, by using topological
arguements, they proved that, although the continuous translation
symmetry is broken in the 2DMM system, the degenerancy of this
zero-energy states is the {\it same} as that for the unmodulated system.
One may wonder whether this unexpected degeneracy is symmetry-related and, 
if it is, what is the nature of this symmetry. In fact, it was shown
by Gendenshtein that, because the Hamiltonian in Eq.~(1) with
$g_e=2$ can be factorized into a product of two conjugate first-order
differential operators, \cite{ac} such a symmetry is indeed present
for the zero-energy states. \cite{susy} However, this is {\it not} a symmetry
of the original Hamiltonian, but rather of one of the two
first-order differential operators. (For more details, see
Ref.~[\onlinecite{susy}].)

In comparison with the $g_e \neq 2$ cases, there is one more unique
feature for the $g_e=2$ case, which becomes obvious when the
modulation is quite strong.  Figs.~4(a) and (b) show the strongly
modulated band structures of {\it spin-down} electrons with $g_e=1$
and 2, respectively.  The modulation strength is
$b_{10}a^2=0.8\phi_0$.  For the case of $g_e=2$, besides the fact that
the lowest energy level remains flat despite the strong mixing between
the unperturbated LLs, it is apparent that the zero-energy level is
isolated from the intermingled fractal structure with a finite energy
gap.  This is true for even stronger modulations.  On the contrary,
such a behavior does not appear in Fig.~4(a) for the case of $g_e=1$.
A simple explanation of the gap above the flat band is as follows:
\cite{dubrovin} if the gap collapses at a particular modulation, such
that a state from a higher energy band merges with the zero-energy
one, then the degenerancy of the zero-energy states will increase by
one.  However, this is impossible because this degenerancy in the
modulated system must be the same as that in the unmodulated one, as
mentioned above.  \cite{dubrovin} Therefore, the flat band has to be
separated from higher bands.

Fig.~5 shows the dependence of the energy bands on the $g$ factor when
the total flux per plaquette is $B_0 a^2=1\phi_0 $ and
$b_{10}a^2=0.2\phi_0$.  In this figure, the sign of the $g$ factor
refers to the direction of spins (+ for up, $-$ for down).
(Note that the sign convention for $g_e$ applies only to Fig.~5, but not
to the previous discussions.) It is
clear that the lowest band is broadened as soon as $g_e \neq -2$.
Notice that the spectrum for $g_e=2$ and that for $g_e=-2$ are
identical, except for the lacking of the flat band at zero energy for
$g_e=2$.  In addition to the zero-energy flat band, the band widths
of other energy bands can also shrink to zero at some particular values of 
$g_e$.
For example, The width of the second lowest energy band is zero when
$g_e=-0.12$.  However, unlike the shrinking of the lowest band at
$g_e=-2$, this `pinch' point moves if a different flux value of $B_0
a^2$ is chosen.

\section{Summary and Discussions}

In this paper, we present an accurate multi-band calculation of the
energy spectra of the 2DMM spin-1/2 electronic systems, in which the
Zeeman effect is also taken into account.  We find that, when the
Zeeman energy is not negligible, the spectra are changed qualitatively
with respect to their electric modulated counterparts.  Moreover, in
the special case when the electron $g$ factor is two, it is found
that:  1.  the positive-energy eigenstates have a two-fold degeneracy
with respect to spin flip; 2.  for the spin-down electrons, the ground
states seem to be unaffected by the periodic modulation and remains
highly degenerate even in very strong modulations; 3.  the ground
states are separated from higher energy states with a finite energy
gap.  However, these special properties no longer exist if the $g$
factor is not equal to two.

In real systems, the electron $g$ factor can be changed by either
varying the width of a quantum well that holds the electrons,
\cite{confine} or by applying a hydrostatic pressure\cite{hydro} to
the sample.  Thus, it is possible to combine these two methods and to
design an experiment in which the $g$ factor can be continuously tuned
around two.\cite{hydro} Under such a circumstance, the peculiar and
robust spectral properties of the lowest energy level should exhibit
itself through the transport properties.  Since the band splitting may
suppress the band conductivity, \cite{2dmmtry} it is advised to keep
the flux per plaquette at one or a simple fraction in order to observe
the conductivity enhancement induced by level broadening away from
$g_e=2$, or, conversely, the conductivity reduction at $g_e=2$.

However, most of current experiments can achieve only weak
modulations.  For example, consider a source of a periodic magnetic
field, $B(\br)=B_0+(B_0/2)(\cos 2\pi x/a +\cos 2\pi y/a)\geq 0$.  The
field modulation felt by a 2DEG at a distance $d$ below the source
becomes $B_d(\br)=B_0 +(B_0e^{-2\pi d/a}/2)(\cos 2\pi x/a +\cos 2\pi
y/a)$.  \cite{rammal} This corresponds to $b_{10}=e^{-2\pi d/a}/4$ in
our calculation.  For typical values such as $a=1\ \mu{\rm m}$ and
$d=10\ {\rm nm}$, $b_{10}$ is equal to 0.23.  The optimum value of
$b_{10}$ is 1/4, when $d=0$.  It can be larger than 1/4 only if the
amplitude of the modulation field is larger than the background field.
In this case, the total field $B(\br)$ reverses direction in some
regions.

In the future, to bring theoretical results much closer to the real
experiments, ingredients such as disorder and electron-electron
interaction have to be included in the calculation.  It may also be
necessary to include an extra two-dimensional electric modulation,
which is inevitably induced due to the strain exerted by the
ferromagnetic or the superconductor grid at low temperatures in recent
experiments.\cite{carmona,ye,2dmm2dem} It will be very interesting to
investigate the influence of these factors on the energy spectra
reported here.

\acknowledgments
This work is supported by the National Science Council of Taiwan under
contract No. 87-2112-M-007-008.

\appendix

\section{Guiding-center Coordinate}
\label{guiding}

In the semiclassical calculations of the transport properties of the
quantum Hall systems, in which the applied magnetic field is strong
and the disorder potential is smooth, \cite{chang97} one usually
decomposes the electron coordinate $\br$ into a fast-moving cyclotron
coordinate, $\bxi={\bf\hat z}\times ({\bf p}+{\bf
A}_0)$, and a slow-moving guiding center,
${\bf R}=\br-\bxi$.  In the present study, although the periodic field
variation is not required to vary smoothly, the derivation of the
multi-band Harper equation can be simplified with the help of this
decomposition.

It can be shown that $(\xi_1,\xi_2)$ and $(R_1,R_2)$ are independent
conjugate pairs respectively, i.e., $[\xi_1,\xi_2]=-i$, $[R_1,R_2]=i$,
and $[\xi_i,R_j]=0$ for $i,j=1,2$.  Thus the exponential
$e^{i\bg\cdot\br}$ can be decomposed as
$e^{i\bg\cdot\bxi}e^{i\bg\cdot{\bf
R}}=e^{i\bg\cdot\bxi}e^{i\bgt\cdot\bRt}$, where $\bgt=\bg\times{\hat
z}$ and ${\bf\tilde R}={\bf R}\times{\bf\hat z}$.  For a Landau gauge 
with ${\bf A}_0(\br)=(-y,0)$, we have $\bxi=(-i\partial_y,i\partial_x+y)$ 
and ${\bf\tilde R}=(-i\partial_x,-i\partial_y-x)$.  Therefore, the magnetic
translation operators in Eq.~(5) can be rewritten as $T_j=e^{i{\tilde
R}_j a_j},\ j=1,2$.
Consequently, we have the following very useful identities,
\bea
T_1|n,k_1\rangle=e^{i{\tilde R}_1 a_1}|n,k_1\rangle &=& e^{ik_1
a_1}|n,k_1\rangle,\cr
T_2|n,k_1\rangle=e^{i{\tilde R}_2 a_2}|n,k_1\rangle &=& |n,k_1-a_2\rangle.
\label{R}
\eea
The second equation is a direct result of the commutation
relation between ${\tilde R}_1$ and ${\tilde R}_2$, $[{\tilde
R}_1,{\tilde R}_2]=i$.  Therefore, the `rotated' guiding-center
coordinate is the generator of the magnetic translation.  It can be
verified that $e^{i{\tilde
R}_1a_1}|n,\bk\rangle=e^{ik_1a_1}|n,\bk\rangle$ and $e^{i{\tilde
R}_2qa_2}|n,\bk\rangle=e^{iqk_2a_2}|n,\bk\rangle$ using Eq.~(\ref{R})
and the periodicity of ${\bar d}_l$ (see Eq.~(\ref{n})).  Thus,
$|n,\bk\rangle$ are indeed the common eigenstates of $H_0$, $T_1$, and
$T_2^q$.  Similarly, due to the periodicity of $d_{n,l}^{(\alpha)}$
(see Eq.~(\ref{E})), one can prove that the energy eigenstates
$|\alpha,\bk\rangle$ are also the eigenstates of $T_1$ and $T_2^q$.

\begin{figure}
%\epsfxsize=3.3truein
%\epsffile{fig1.eps}
%\vspace{.8 cm}
\caption{The lowest five energy bands for spin-down electrons with
$g_e=0$ under a weak square modulation field.  The modulation strength
is fixed at $b_{10}a^2=0.2\phi_0$ while the total flux per plaquette,
$B_0 a^2$, varies.  The energy is in units of $\hbar\omega_c$ and the
inverse flux per plaquette is in units of $\phi_0^{-1}$.}
\end{figure}

\begin{figure}
%\epsffile{fig2a.eps}
%\epsffile{fig2b.eps}
\caption{The lowest five energy bands for a spin-down (a) and
a spin-up (b) electrons with $g_e=1$. The modulation strength
and the units being used are the same as those in Fig.~1.}
\end{figure}

\begin{figure}
%\epsffile{fig3a.eps}
%\epsffile{fig3b.eps}
\caption{The lowest five energy bands for a spin-down (a) and a
spin-up (b) electrons with $g_e=2$.  The modulation strength and the
units being used are the same as those in Fig.~1.  It can be seen that
the lowest energy level for a spin-down electron is flat.  Also, the
spectrum is degenerate with respect to spin flip, except for the
zero-energy flat band in (a).}
\end{figure}

\begin{figure}
%\epsffile{fig4a.eps}
%\epsffile{fig4b.eps}
\caption{The lowest five energy bands for a spin-down electron with
$g_e=1$ (a) and 2 (b). The modulation strength is $b_{10}a^2=0.8\phi_0$.}
\end{figure}

\begin{figure}
%\epsffile{fig5.eps}
\caption{The dependence of the band widths on the $g$
factor when the total flux per plaquette $B_0 a^2=1\phi_0$
and $b_{10}a^2=0.2\phi_0$. The sign of the $g$ factor refers to
the direction of spins ($+$ for up, $-$ for down). The energy bands at 
$g_e=-2$ are indicated by vertical bold lines.} \end{figure}

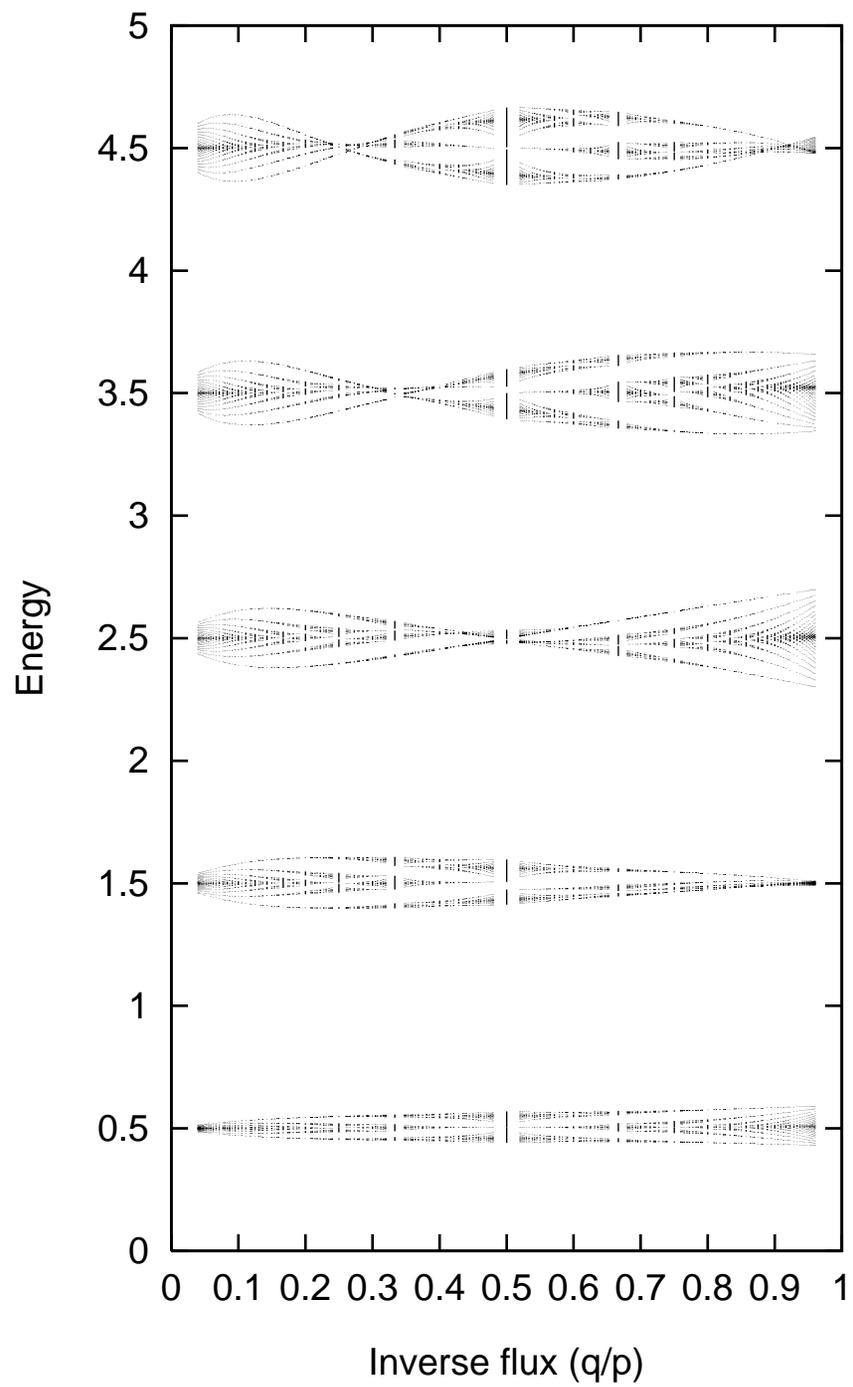
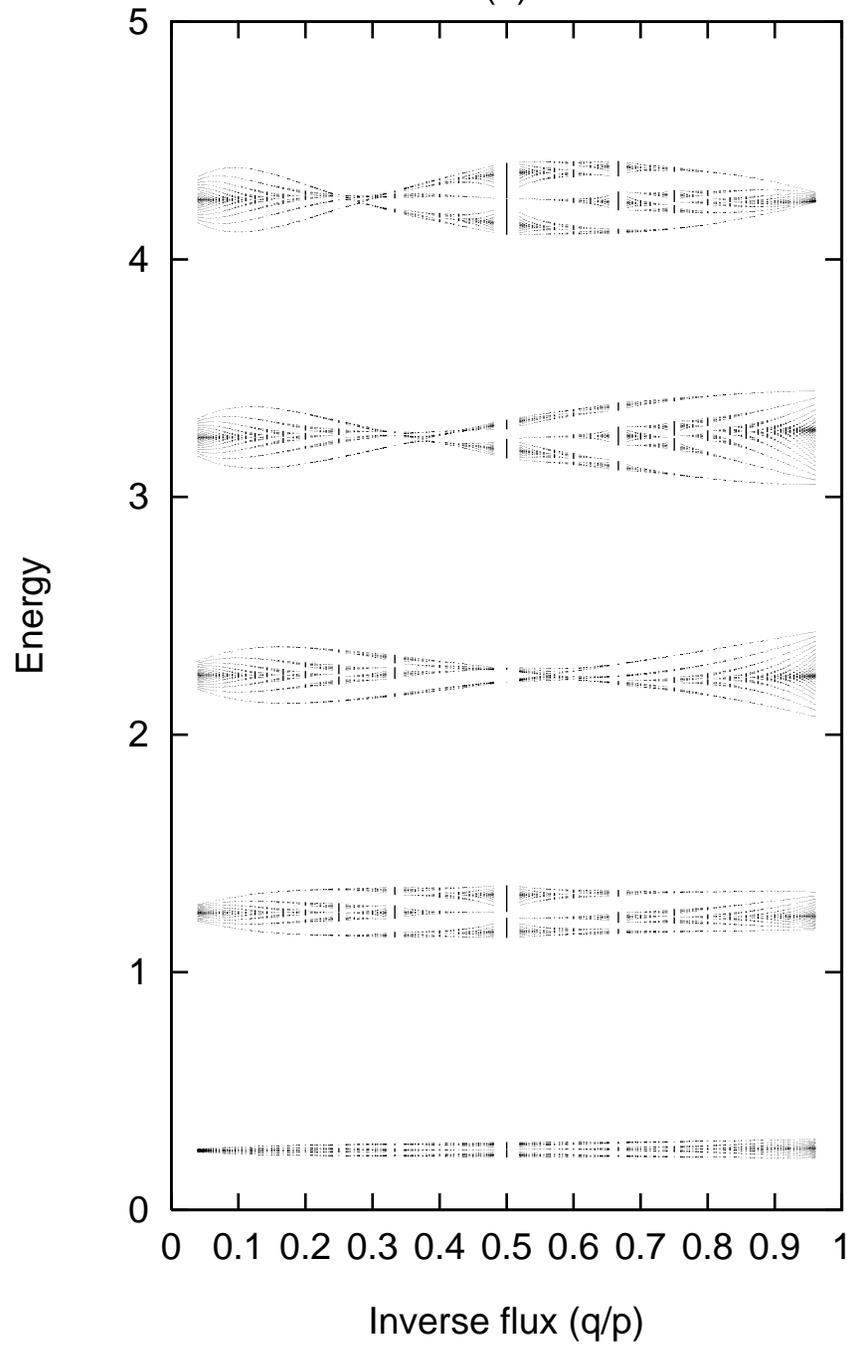
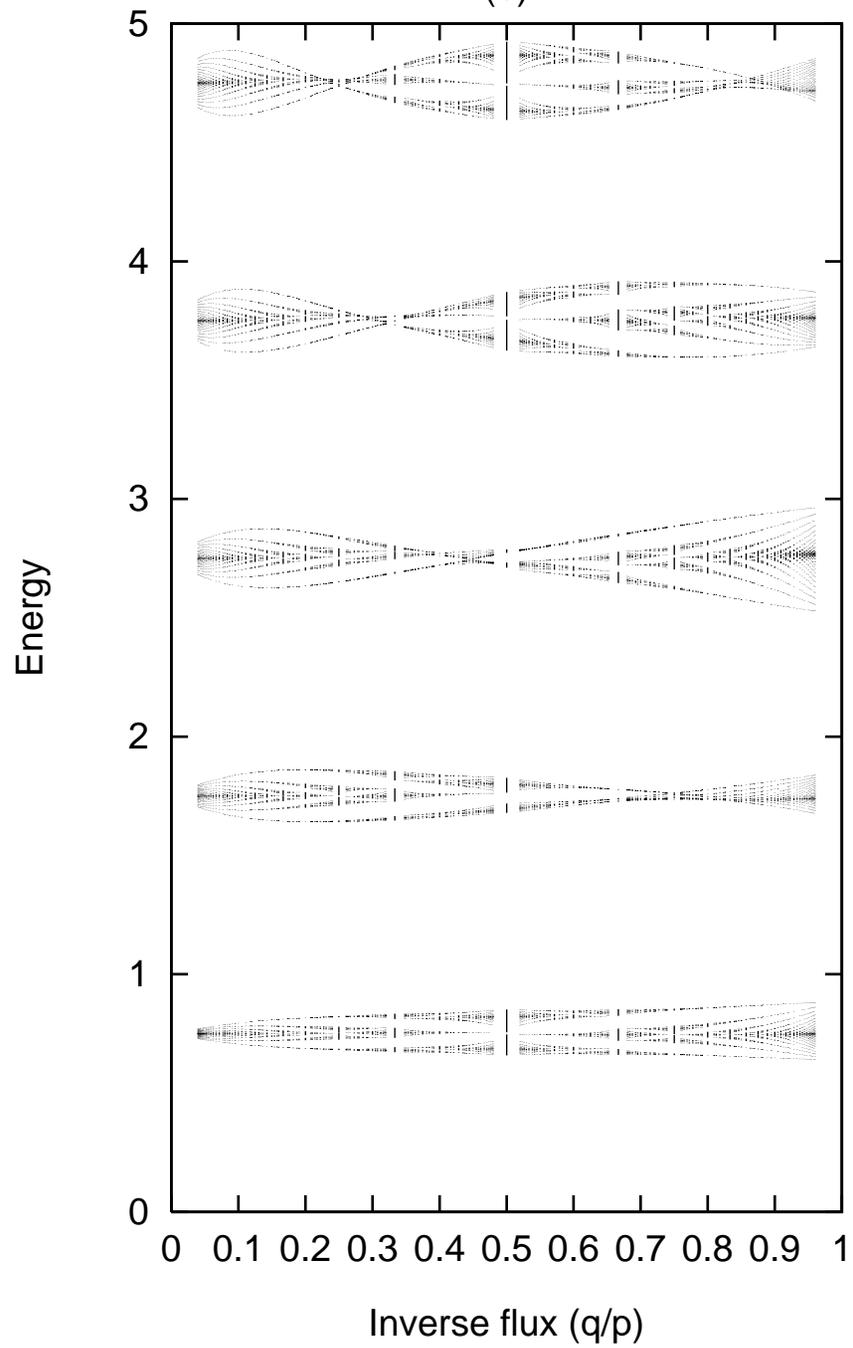
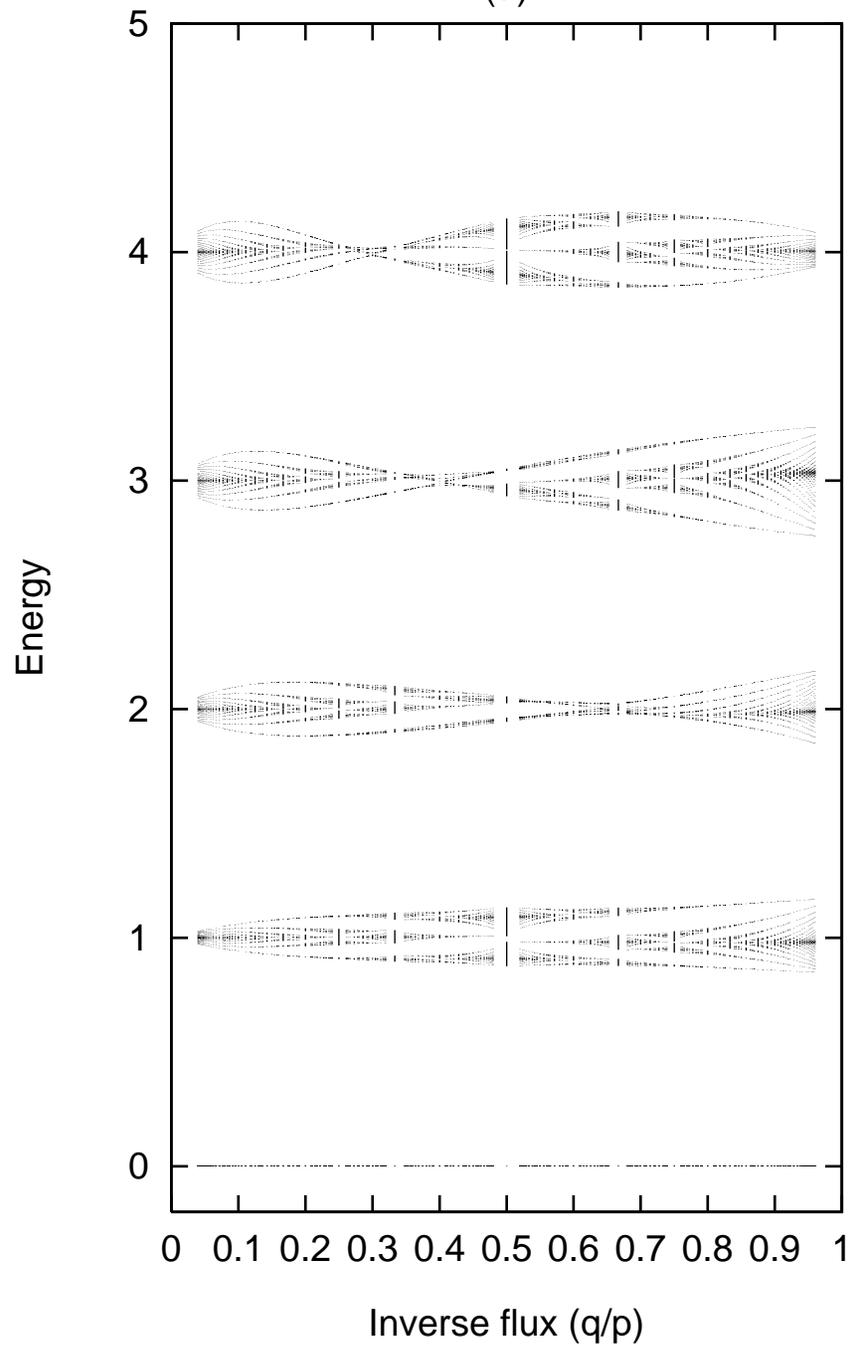
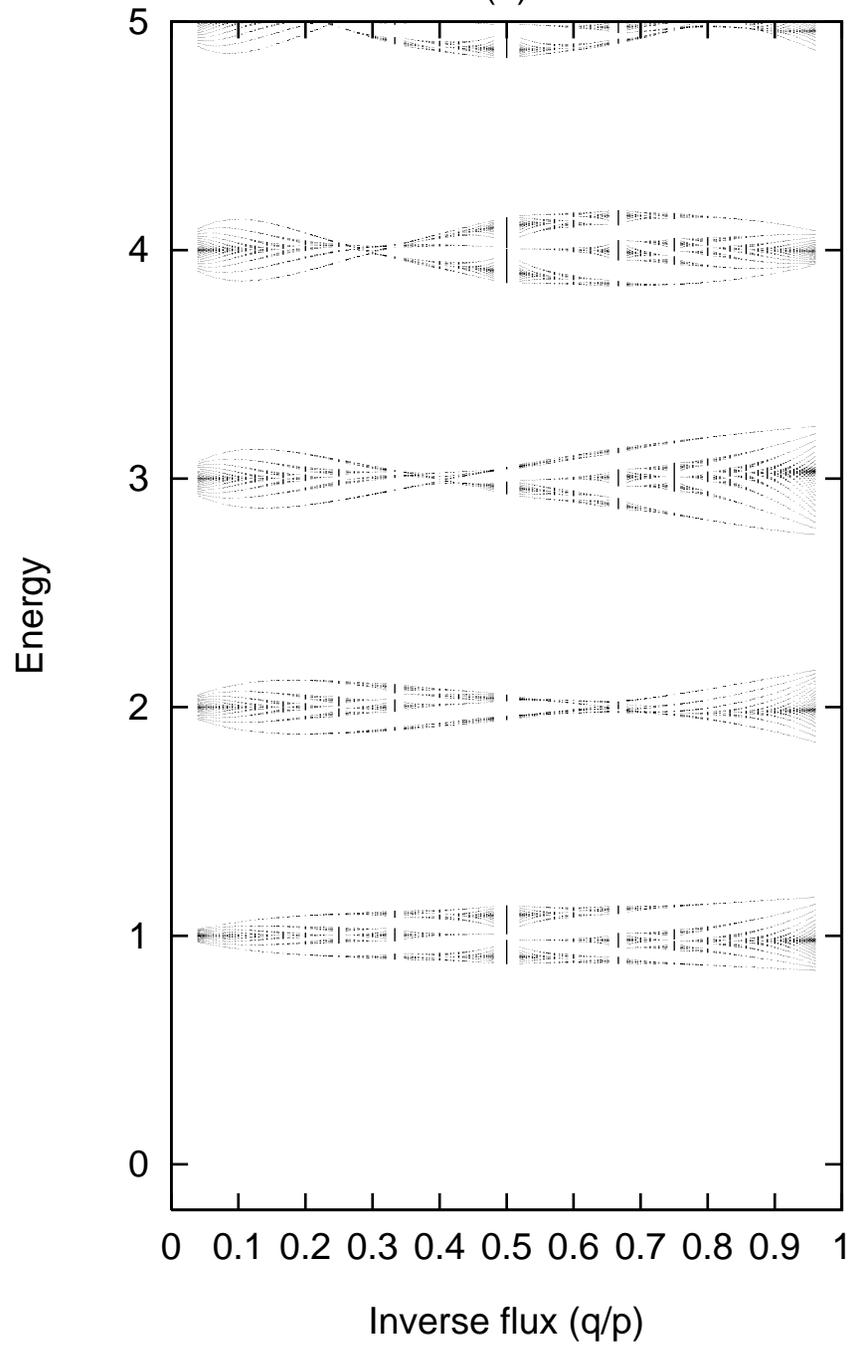
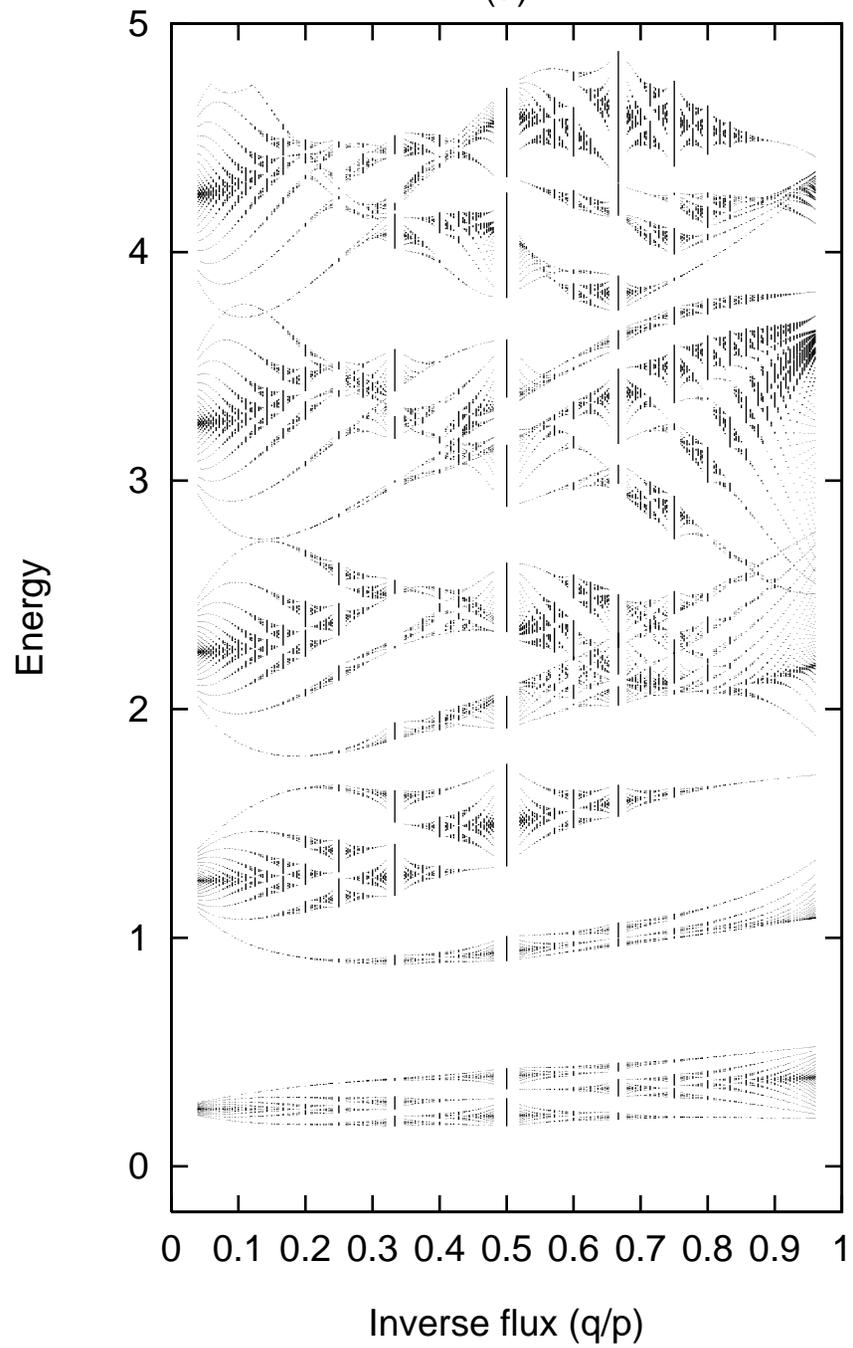
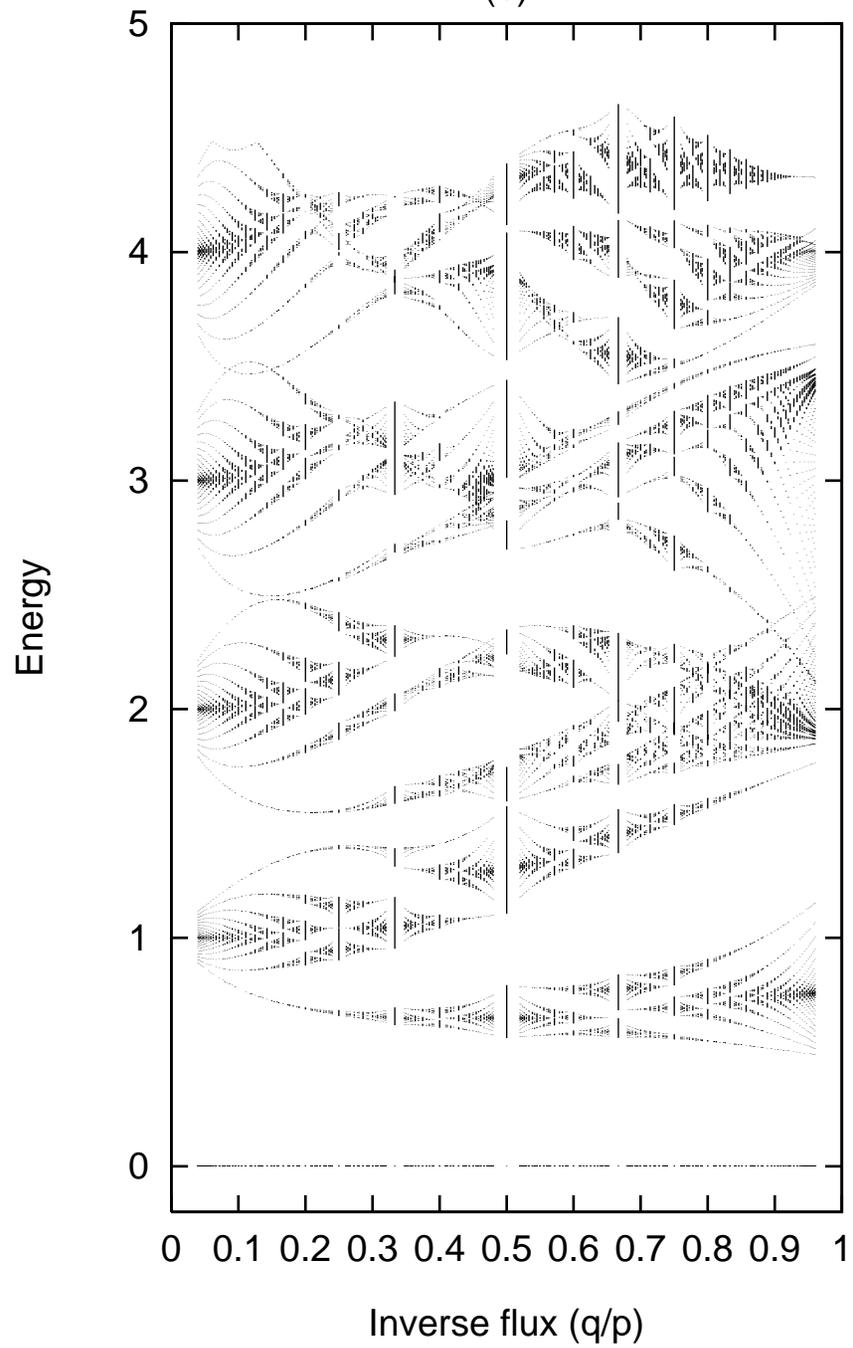
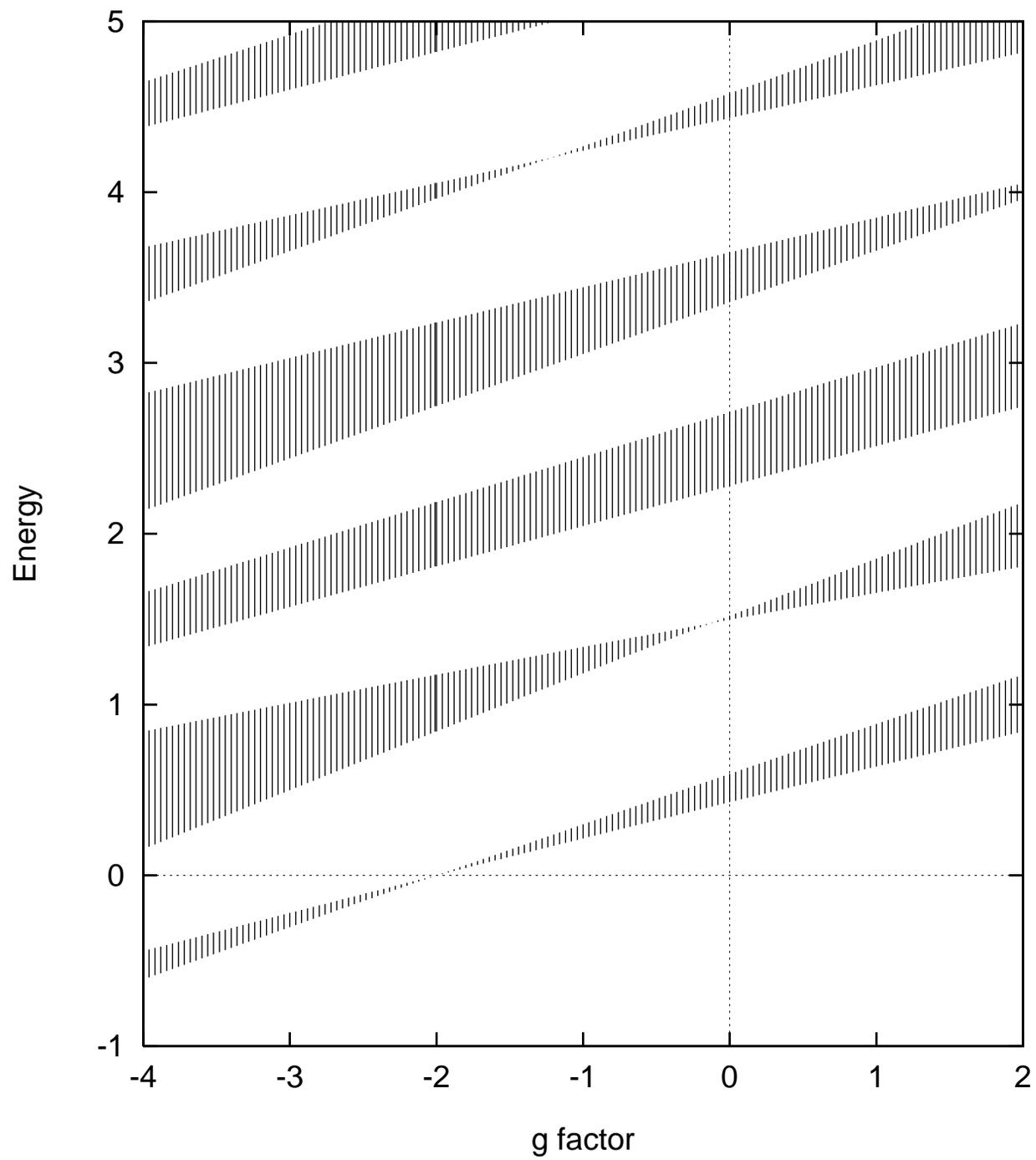
\end{document}